\documentclass[11pt]{article}
\usepackage{blois,epsfig}

\usepackage{amsmath,amssymb}

\def\Journal#1#2#3#4{{#1} {\bf #2}, #3 (#4)}

\def\PRL{\em Phys. Rev. Lett.}
\def\PRD{{\em Phys. Rev.} D}

\def\be{\begin{equation}}
\def\ee{\end{equation}}

\newcommand{\mtop}{\ensuremath{m_{\text{top}}}}
\newcommand{\mW}{\ensuremath{m_{\text{W}}}}
\newcommand{\mZ}{\ensuremath{m_{\text{Z}}}}
\newcommand{\mHiggs}{\ensuremath{m_{\text{H}}}}
\newcommand{\ppbar}{\ensuremath{\text{p}\bar{\text{p}}}}
\newcommand{\invpb}{\ensuremath{\ \text{pb}^{-1}}}
\newcommand{\invfb}{\ensuremath{\ \text{fb}^{-1}}}
\newcommand{\ra}{\ensuremath{\rightarrow}}
\newcommand{\MET}{\mbox{\ensuremath{\,\slash\kern-.7em E_{\text{T}}}}}
\newcommand{\pt}{\ensuremath{p_{\text{T}}}}
\newcommand{\ut}{\ensuremath{u_{\text{T}}}}
\newcommand{\vut}{\ensuremath{\vec{u}_{\text{T}}}}
\newcommand{\mt}{\ensuremath{m_{\text{T}}}}
\newcommand{\epem}{\ensuremath{\text{e}^+\text{e}^-}}
\newcommand{\Jpsi}{\ensuremath{\text{J}/\psi}}

\begin{document}
\vspace*{4cm}
\title{MEASURING THE W BOSON MASS AT THE TEVATRON}

\author{ M.P.~SANDERS }

\address{Ludwig-Maximilians-Universit\"at, Fakult\"at f\"ur Physik, \\
Am Coulombwall 1, D-85748 Garching, Germany}

\author{ For the D0 and CDF Collaborations }
\address{}

\maketitle\abstracts{
The measurement of the mass of the W boson is one of the prime
goals of the Tevatron experiments. In this contribution, a
review is given of the most recent determinations of the W boson mass (\mW{})
at the Tevatron. The combined Tevatron result, $\mW = 80.420 \pm
0.031$~GeV, is now more precise than the combined LEP result, leading
to a world average value of $\mW = 80.399 \pm 0.023$~GeV. 
}

\section{Introduction}

The mass of the W boson (\mW{}) is a key parameter in  the
electroweak model. It is connected to other parameters of the model
through the following equation:
\be
\mW^{\text{2}}\left( 1-\frac{\mW^{\text{2}}}{\mZ^{\text{2}}} \right) = 
   \frac{\pi\alpha}{\sqrt{\text{2}}G_{\text{F}}} \left(
   \frac{1}{1-\Delta r} \right),  
\ee
with \mZ{} the mass of the Z boson, $\alpha$ the fine-structure
constant and $G_{\text{F}}$ the Fermi 
coupling constant. The term $\Delta r$
describes higher-order
corrections which depend quadratically on the top quark mass and
logarithmically on the Higgs boson mass. Precise measurements of the
top quark mass and the W boson mass can thus provide a prediction for
the mass of the undiscovered Higgs boson. For the top quark mass and
the W boson mass to have equal weight in this prediction, 
the W boson mass has to be determined very precisely: 
$\Delta\mW \approx
\text{0.006} \cdot \Delta\mtop$. With 
$\Delta\mtop = 1.1$~GeV~\cite{ShabalinaBlois2010}, 
$\Delta\mW$ would have to be 7~MeV.

The measurements of the W boson mass 
presented here were performed by the D0 and CDF experiments at the Tevatron, 
a \ppbar{} collider at a centre-of-mass energy of
1.96~TeV. 
These results are based on an integrated luminosity of 1\invfb{} for
D0~\cite{D0_pub}
and 200\invpb{} for CDF~\cite{CDF_pub}.
The total integrated luminosity
delivered to each of the experiments is more than 9\invfb{}. 

\section{Measurement Strategy}

In hadron collisions, only the W\ra{}e$\nu$ and W\ra{}$\mu\nu$ decays
can be used for precise mass measurements; hadronic W boson decays
cannot be distinguished from the large multi-jet background and the 
W$\ra\tau\nu$ decays cannot be measured precisely enough. The
experimental event
signature is then rather simple: a lepton and a neutrino with large
momentum transverse to the beam direction (\pt{}). The neutrino can be
identified as missing 
transverse energy (\MET{}). The neutrino momentum in the direction of
the colliding beams cannot be determined. 

In the D0 analysis, both $\pt(\ell)$ and
\MET{} are required to be at least 25~GeV, in the CDF analysis the cut
is at 30~GeV. The lepton has to be well identified in the barrel
portion of the detector, $|\eta| < 1$ ($\eta$ is the pseudorapidity). 
In addition, the transverse recoil momentum of the W
boson (\ut{}) is required to be small, $\ut < 15$~GeV.
Leptonic decays of the Z boson (Z$\ra\ell\ell$) are selected in a
similar fashion, and serve as a calibration sample.
The D0 collaboration analyzes the electron channel only, whereas the
CDF collaboration uses both electrons and 
muons.

The W boson mass is 
extracted from fits to distributions of kinematic quantities,
simulated for a range of hypothetical W boson masses (template fits).
The quantities considered are the transverse momentum of the
lepton, the transverse momentum of the neutrino 
$p_{\text{T}}(\nu) = \MET$ with
$\vec{p}_{\text{T}}(\nu) =
-\vec{p}_{\text{T}}(\ell) - \vut$,
and the transverse
mass of the lepton-neutrino system. 
The transverse mass is defined by 
$ \mt^{\text{2}} = \text{2}
\pt(\ell)\pt(\nu)(\text{1}-\cos\Delta\phi_{\ell\nu})$, where
$\Delta\phi_{\ell\nu}$ is the difference in azimuthal angle between
the lepton and the neutrino. 

To get a precise W boson mass measurement, the challenge is to
accurately calibrate the measurable quantities, $\pt(\ell)$ and \ut{},
and to model the W boson production and the detector in a
parameterized simulation. This simulation is then used
to produce the fit templates. 

\subsection{Production}

Production of W bosons is
modeled with the \textsc{resbos} event generator 
and
photon radiation is described with \textsc{photos}
or \textsc{wgrad}. Uncertainties in these theoretical models or the
parameters used, and uncertainties on parton distribution functions (PDFs)
lead to systematic uncertainties on the W boson mass. Both
D0 and CDF estimate an uncertainty on the W boson mass of 10~MeV due to the PDF
uncertainty, and 7~MeV (D0) to 11~MeV (CDF) due to photon radiation. 

\subsection{Lepton Momentum Calibration}

The D0 measurement relies on a precise calorimeter calibration. The
amount of energy that an electron loses in uninstrumented material is
corrected for using a detailed \textsc{geant} simulation of the
calorimeter, in combination with measured energy depositions in the
four electromagnetic layers of the calorimeter, as a function of
energy. This leads to energy and $\eta$ dependent parameterizations of
the response and resolution of the calorimeter. The electron momentum
scale and offset are then set using a data sample of 18700 Z$\ra\epem$
events and the world average value of the Z boson mass.

In the CDF analysis, the central tracking detector is used to calibrate the
lepton momentum. 
The elements of the tracking detector are aligned with cosmic muons.
A sample of 600000 $\Jpsi\ra\mu^+\mu^-$ decays
is then used to determine the momentum scale. By doing this in bins of 
$1/\pt(\mu)$ and $\cot\theta_\mu$, the description of the material in
the detector and of
the non-uniformities in the magnetic field can be fine-tuned. The electron
momentum scale is 
set by comparing the energy measurement in the calorimeter with
the momentum measurement in the tracking detector. 

Uncertainties on the lepton momentum scale are the dominant sources of
systematic uncertainty on the W boson mass. For D0,
a 34~MeV uncertainty is assigned. CDF assigns a 17~MeV uncertainty in
the muon channel and 30~MeV in the electron channel. These
uncertainties are largely of statistical nature and will decrease with
larger calibration samples.

\subsection{Recoil Momentum Calibration}

The  recoil \vut{} consists not only of a ``hard'' component due
to the particles that are truly recoiling against the W boson, but also of
``soft'' components which cannot be distinguished from the hard component on an
event-by-event basis. These are, e.g., contributions due to
the underlying event or additional hadron-hadron interactions.
The hard and soft components of the recoil are
modeled separately. A sample of Z boson decays is then used to fix the
recoil momentum scale and resolution, using the imbalance between the transverse
momentum of the Z boson (determined using the well-measured leptons)
and the measured recoil momentum.

The magnitude of the systematic uncertainty on the W boson mass due to
the recoil modeling strongly depends on which kinematic variable is used to
fit for the W boson mass. For D0 (CDF), the
uncertainties are 6~MeV (9~MeV) for the fit using the transverse mass, 12~MeV
(17~MeV) for the fit with the lepton transverse momentum and 20~MeV (30~MeV) for
the fit to the \MET{} spectrum.

\section{Results}

\begin{figure}[b]
\begin{center}
\psfig{figure=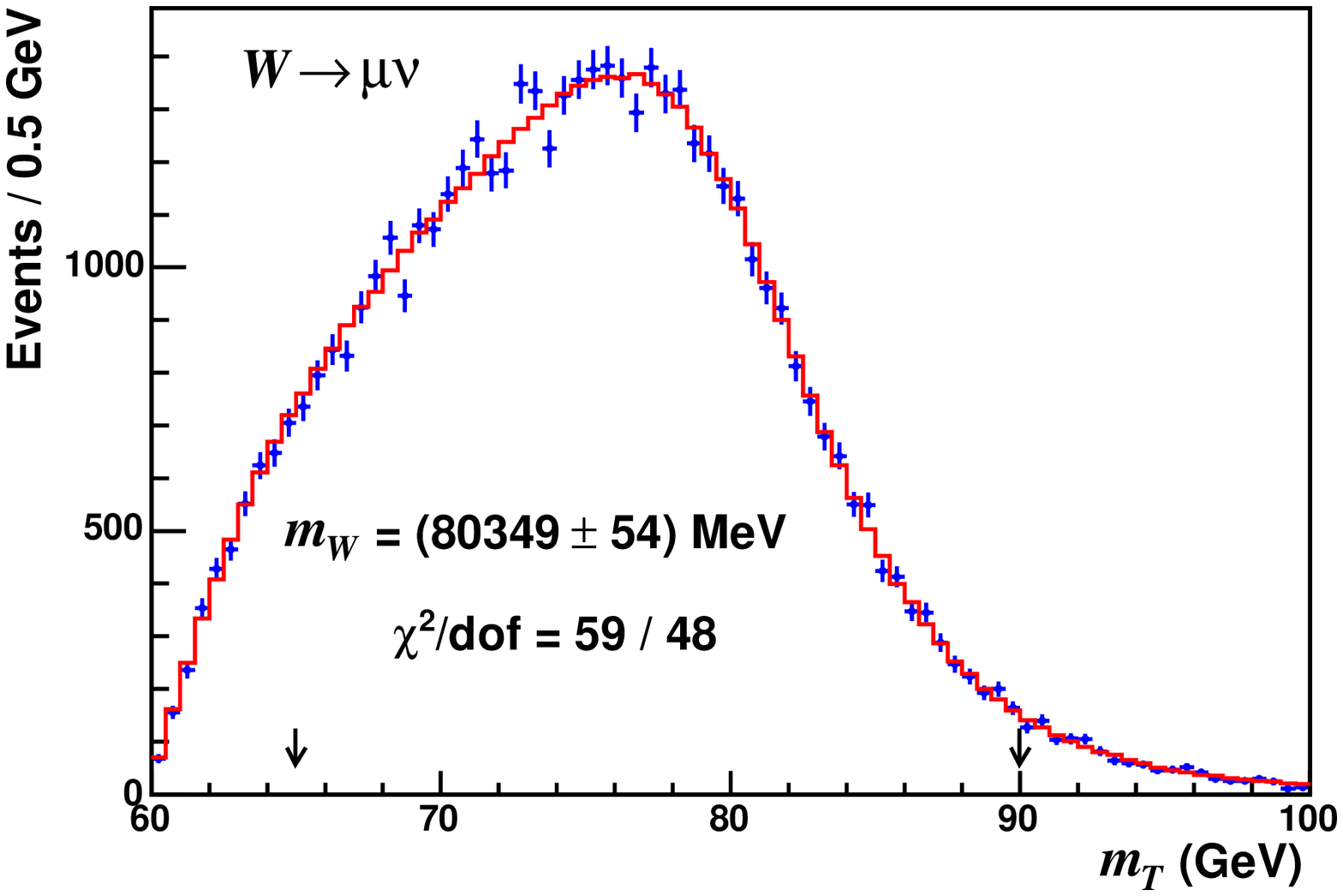,width=6cm}%
\psfig{figure=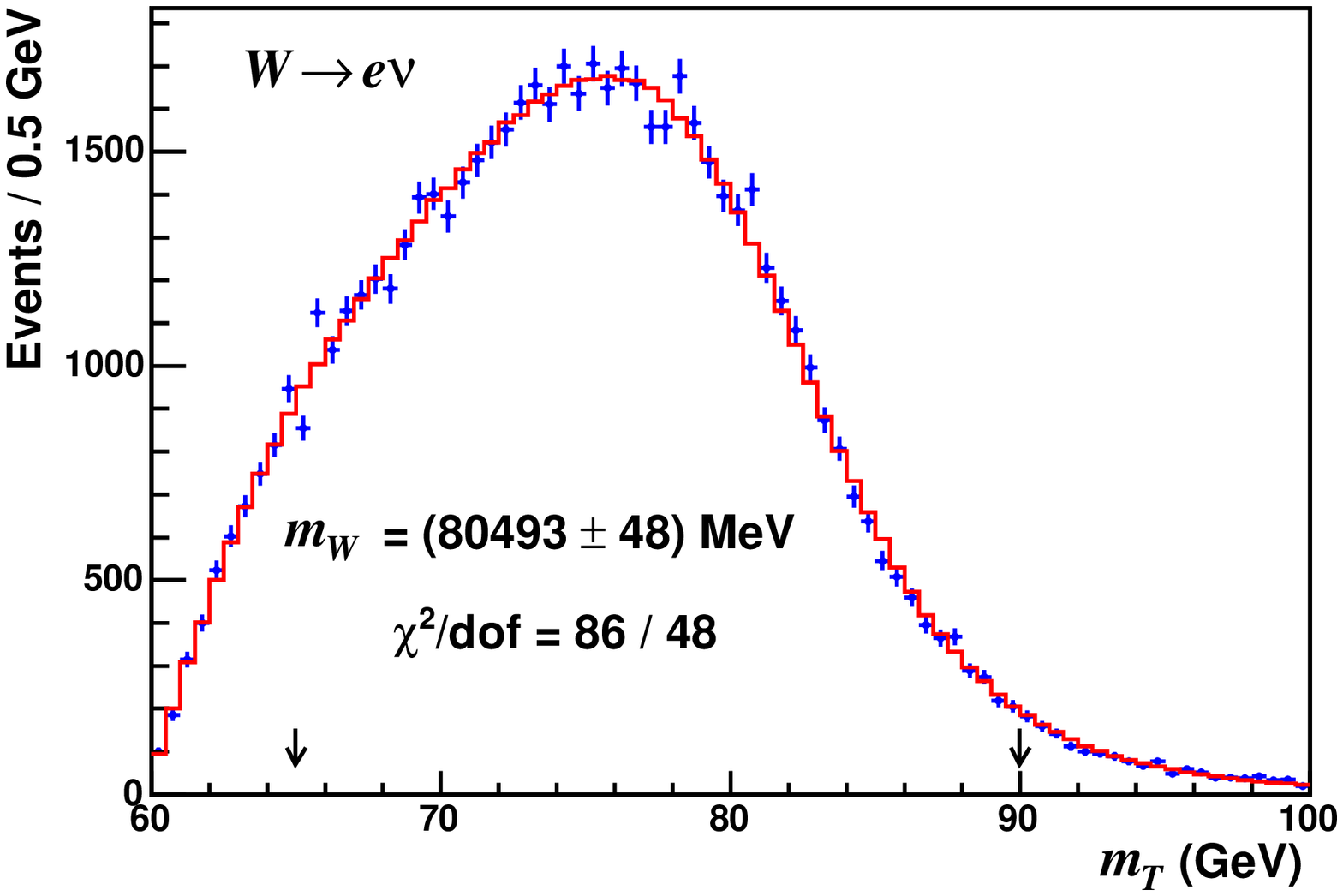,width=6cm}
\caption{Distributions of the transverse mass in the muon channel
  (left) and electron channel (right) in 200\invpb{} of CDF data. The data
  are shown by the points, the histogram indicates the simulation for
  the best-fit \mW{}. The fit range is indicated by the arrows.}
\label{fig:mWCDF}
\end{center}
\end{figure}

The CDF collaboration analyzed a data-set corresponding to an
integrated luminosity of 200\invpb{}. In the muon (electron) channel, 
51000 (64000) W boson
candidates were identified. 
In Fig.~\ref{fig:mWCDF}, distributions of the transverse
mass in the two channels are shown. The data points agree well
with the expectation from the parameterized simulation, for the
best-fit value of the W boson mass. 
A statistical combination, taking
into account all correlations, of in total six measurements (electron
and muon channel, using $\pt(\ell)$, \MET{} or \mt{}) yields 
$\mW = 80.413 \pm 0.034_{\text{stat}} \pm
0.034_{\text{syst}}$~GeV, corresponding to a total uncertainty of 48~MeV.

\begin{figure}[tb]
\begin{center}
\psfig{figure=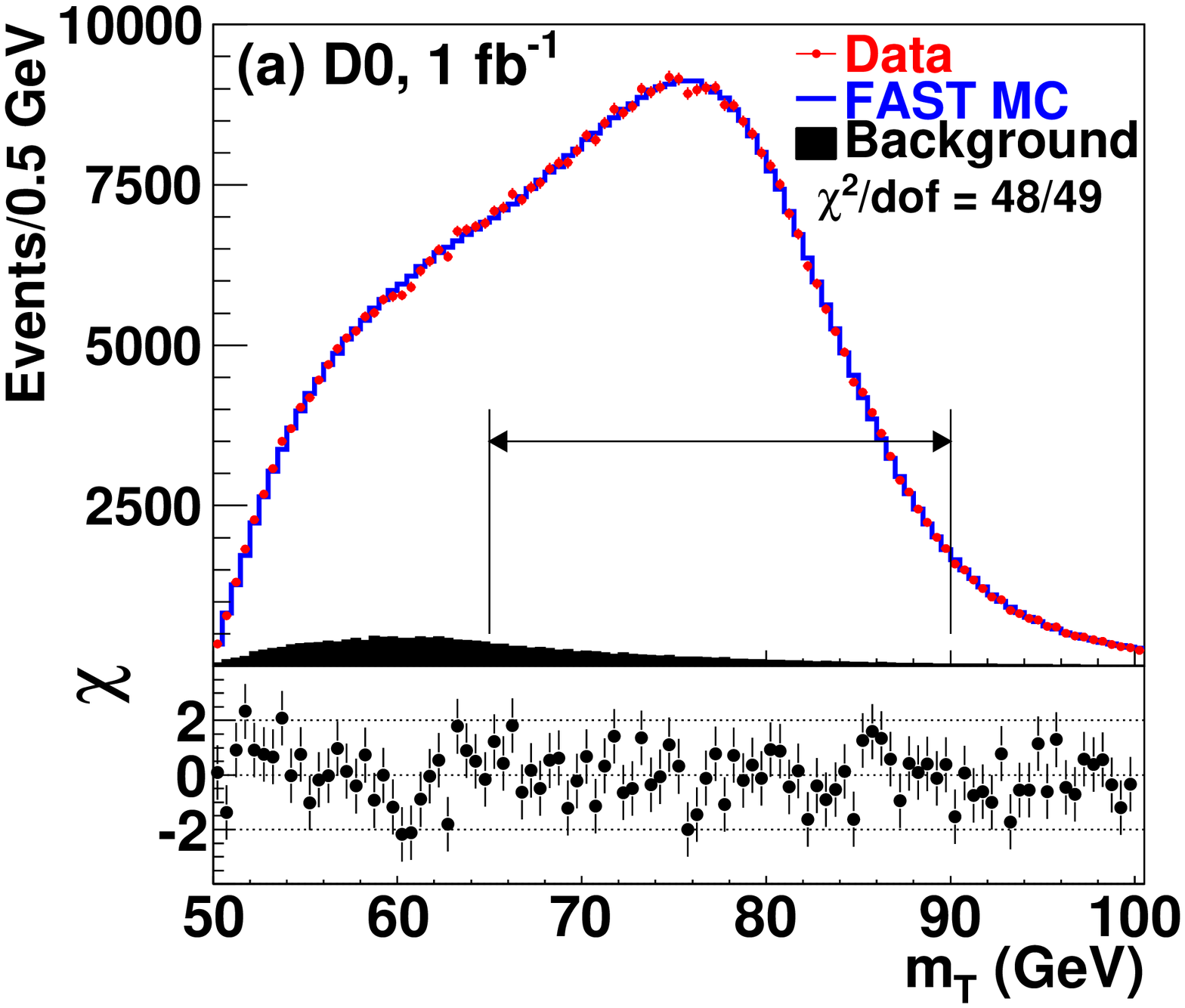,width=5.3cm}%
\psfig{figure=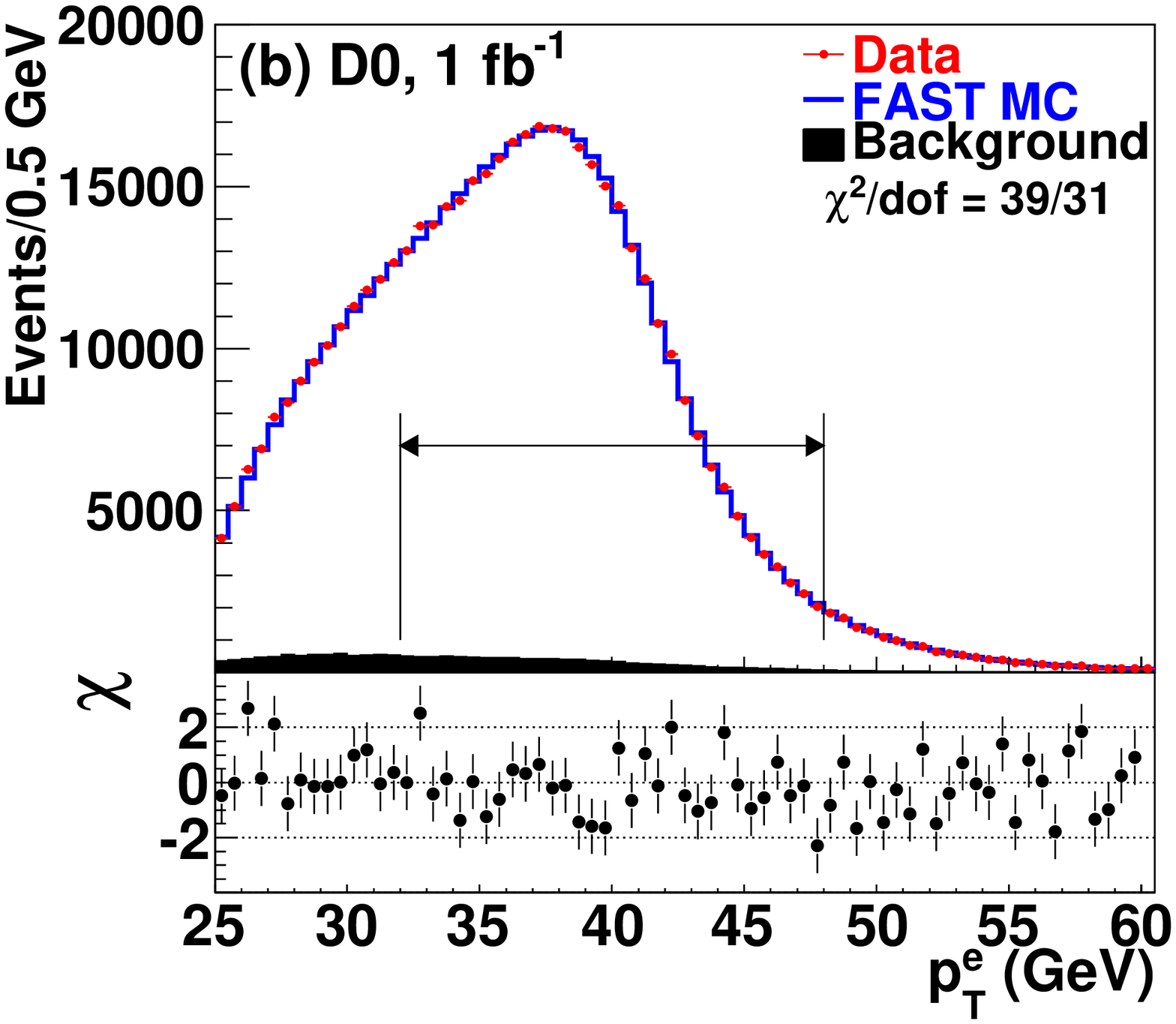,width=5.3cm}%
\psfig{figure=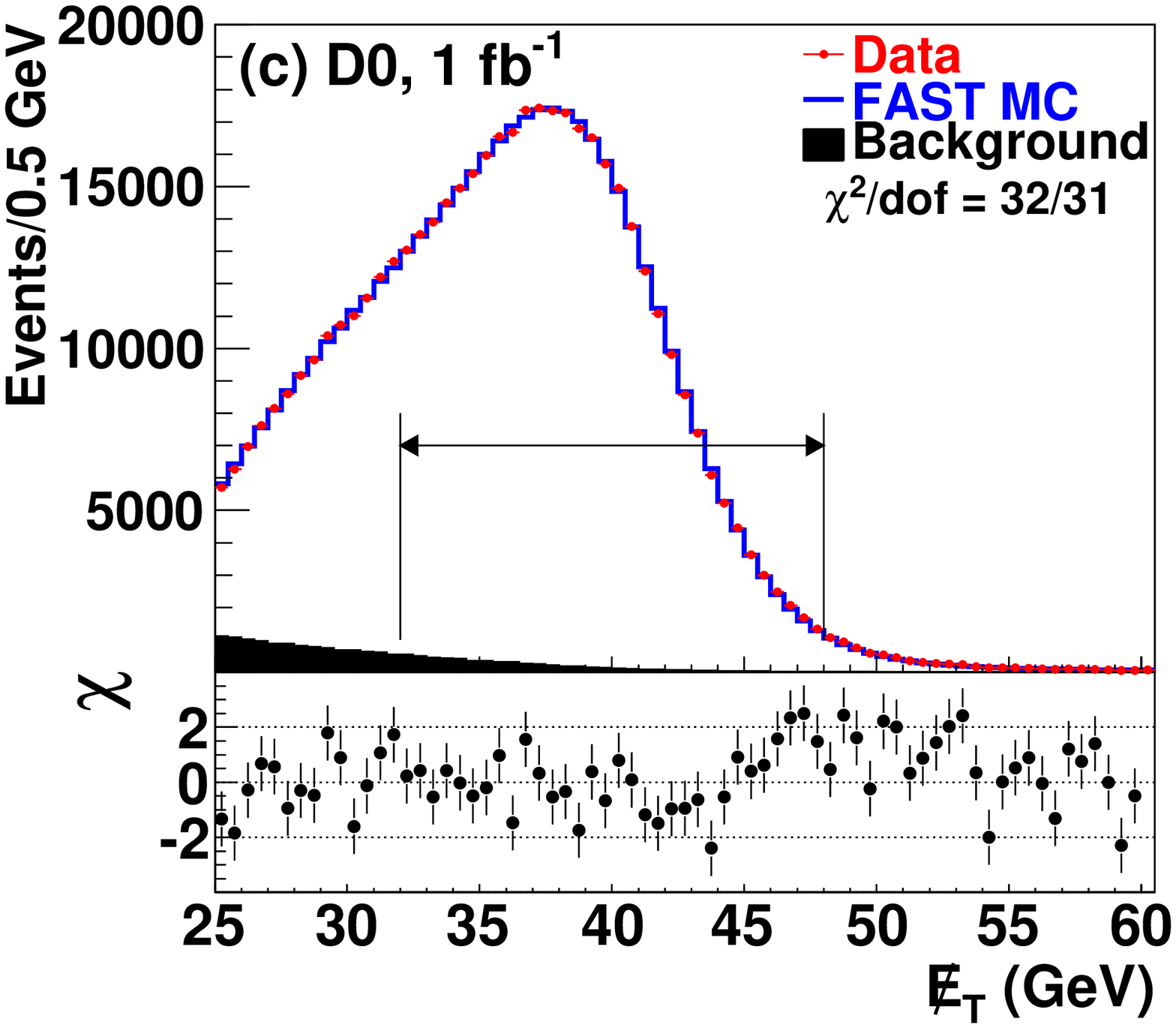,width=5.3cm}
\caption{Distributions of the transverse mass (left), electron
  transverse momentum (middle) and missing transverse energy (right)
  in 1\invfb{} of D0 data. The data
  are shown by the points, the histogram indicates the simulation for
  the best-fit \mW{}. The fit range is indicated by the arrows. The
  $\chi$ values shown below each histogram are the differences between
  data and simulation, divided by the statistical uncertainty.}
\label{fig:mWD0}
\end{center}
\end{figure}

The results of the D0 analysis, based on a data-set corresponding to
an integrated luminosity of 1\invfb{}, are shown in
Fig.~\ref{fig:mWD0}.  Very good agreement is seen
between the 500000 W boson candidates in data and the simulation. The 
three measurements of \mW{} are: 
$80.401 \pm 0.023_{\text{stat}}$~GeV for the fit to the transverse mass, 
$80.400 \pm 0.027_{\text{stat}}$~GeV for the lepton transverse momentum and 
$80.402 \pm 0.023_{\text{stat}}$~GeV for the neutrino transverse
momentum.
Combining these three measurements,
taking into account the strong correlations between them gives $\mW =
80.401 \pm 0.021_{\text{stat}} \pm 0.038_{\text{syst}}$~GeV, or a
total uncertainty of 43~MeV. This is the most precise single
measurement of the W boson mass.  

\subsection{Combination}

The two W boson mass measurements presented here are in excellent agreement
with each other. These measurements and previous Tevatron
measurements have been combined, taking into account correlated
systematic uncertainties (due to parton distribution functions, photon
radiation and the W boson width). The combination  
also corrects the measurements to
the same value of the W boson width. The result,
$\mW = 80.420 \pm 0.031$~GeV, is in good agreement with the LEP
combined measurement of $\mW = 80.376 \pm 0.033$~GeV.
Based on the Tevatron and LEP measurements, the world average is $\mW
= 80.399 \pm 0.023$~GeV. 

\subsection{Interpretation}

As mentioned in the introduction, direct measurements of the W boson
mass and the top quark mass (and other parameters) constrain the mass
of the Higgs boson, assuming validity of the electroweak model. With
the W boson mass measurement presented here, and $\mtop = 173.1 \pm
1.3$~GeV, the Higgs boson mass is found to be $\mHiggs =
87^{+35}_{-26}$~GeV~\cite{LEPEWWG}.

After this conference, a new top quark mass
measurement~\cite{ShabalinaBlois2010} was included
in the standard model fit~\cite{LEPEWWG}. 
This leads to a shift in the expected Higgs
boson mass of $+2$~GeV, but no reduction in the uncertainty. 

\section{Conclusion and Outlook}

The world's most precise W boson mass measurements are now obtained by
the Tevatron experiments: $\mW = 80.420 \pm 0.031$~GeV. Combining
these results with those from LEP leads to an average of
$\mW = 80.399 \pm 0.023$~GeV. 

The precision of the Tevatron measurements is expected to improve
significantly in the future.  The addition of more data will 
decrease both the statistical and systematic 
uncertainties, since the estimation of systematic uncertainties is
largely based on data samples, and therefore also limited by the size
of the data set.

The CDF collaboration has already performed preliminary studies
of a larger data
sample of up to 2.4\invfb{}. 
The additional data were recorded at a
higher instantaneous luminosity, which implies that some deterioration
in, e.g., the recoil resolution is to be expected and to be accounted for.
No significant degradation of the transverse mass resolution was
found~\cite{CDF_update}. 

The final precision of the Tevatron experiments on the W boson mass
could reach the level of 15~MeV \cite{StarkKotwal}, which will also
significantly improve the prediction of the Higgs boson mass.

\end{document}